\documentclass[12pt]{article}
\usepackage{graphicx}
\usepackage{mathptmx,mathrsfs}
\usepackage{amsfonts,amsmath,amssymb,amsthm}
\usepackage{indentfirst}
\usepackage{cite}
\usepackage{hyperref}

\numberwithin{equation}{section}
\def\sumint{\sum \!\!\!\!\!\!\!\! \int }

\begin{document}

\title{Thermal effective Lagrangian of generalized electrodynamics in static
gravitational fields}
\author{R. Bufalo$^{1,2}$\thanks{%
rodrigo.bufalo@helsinki.fi} \\
\textit{{$^{1}${\small Department of Physics, University of Helsinki, P.O.
Box 64}}}\\
\textit{\small FI-00014 Helsinki, Finland}\\
\textit{{$^{2}${\small Instituto de F\'{\i}sica Te\'orica (IFT), Universidade Estadual Paulista}}} \\
\textit{\small Rua Dr. Bento Teobaldo Ferraz 271, Bloco II, 01140-070 S\~ao Paulo, SP, Brazil}\\
}
\maketitle
\date{}

\begin{abstract}
In this paper we compute the effective Lagrangian of static gravitational fields interacting with thermal fields of generalized electrodynamics at high temperature. We employ the usual Matsubara imaginary-time formalism to obtain a closed form expression to the thermal effective Lagrangian at one-loop and two-loop order, in an arbitrary $\omega$-dimensional spacetime, in which the equivalence between the static hard thermal loops and those with zero external energy-momentum is widely explored. Afterwards, the symmetries of the resulting expressions are discussed as well as the presence of the Tolman \emph{local} temperature.
\end{abstract}

\newpage

\tableofcontents

\section{Introduction}

\label{intro}
Thermal field theory is the appropriate framework in describing models in thermodynamical equilibrium, as well as to address the question of stability of these models as a function of environmental variations. In order to have a better understanding in the model's behavior within several regimes depending at the system's temperature, much attention has been devoted in studying the high-temperature limit of the Green's functions, in general, hard thermal loop amplitudes. It then came to attention that in order to have a proper resummation procedure, in which is necessary to control infrared divergences and give physical meaningful outcomes in perturbation theory, one needs to take into account the \emph{hard thermal loops} \cite{ref1}. Basically, this corresponds, in the momentum space, to consider that all the external momenta and energies of the thermal amplitude are much smaller than the temperature $T$.

There are some special cases in field theories, namely, the static and the long wavelength limits, in which it is known that the thermal amplitudes are local functions of the external fields and possess, in general, a closed form expression. In particular, in the case of gauge theories, it has been shown that it is possible to construct closed form expressions for the effective Lagrangian, which generates all thermal Green's functions\cite{ref1,ref2,ref3,ref4}.

One may actually say that the symmetries of a given system are one of the most important characteristic which are widely explored in several branches within theoretical physics. For instance, in gauge theories, there are Ward identities which relate the thermal amplitudes with each other. Not that different, there are similar quantities when one employs the approach of hard thermal loops in a background of soft gravitational fields, where the gravitational thermal amplitudes satisfy simple Ward identities due to the local coordinate transformations \cite{ref5}. Furthermore, the aforementioned limits, static and the long wavelength, are also present in the evaluation of expressions for the effective Lagrangian, but translated into the following form: when the background gravitational field is either time or space independent, respectively. However, the frail point of these limits are that each of them yields two different local effective Lagrangian functionals \cite{ref6,ref7,ref8}.

Nevertheless, recently, an interesting and rich result has been obtained, in which it is shown at one-loop \cite{ref9} and subsequently at two-loop \cite{ref10} calculations, that a static background is equivalent to a spacetime independent configuration, in the high-temperature limit. More precisely, it has been shown that the static limit of thermal Green's functions in gauge theories coincide with the limit when all the external four-momenta are set to zero. Lately, these results were generalized to all orders \cite{ref11}. Another example of the usefulness of this method may also be found in the discussion on thermal scalar fields in a static background \cite{ref12}, which derives in a much more simple manner a previous known result \cite{ref1}.

These aforementioned results together, especially the equivalence between the static hard thermal loops and zero external energy-momentum, have lead to a rather interesting study of a QED plasma in a background of static gravitational fields \cite{ref13}. Besides the rich symmetries of the system, a closed form expression was evaluated for the effective Lagrangian up to two-loop order. Also, remarkably, it was shown that the resulting expression is equivalent to the pressure of a QED plasma in Minkowski spacetime, with the global temperature replaced by the Tolman local temperature \cite{ref14}. This behavior is in agreement with the so-called Tolman-Ehrenfest effect, which states that a system at thermal equilibrium in a stationary gravitational fields, has its temperature varying with the spacetime metric and it is characterized by: $T_{loc}\equiv \frac{T}{\sqrt{g_{00}}}$.

As it has been pointed out in several works \cite{ref15,ref16,ref17,ref18,ref19,ref20} along the years, it is long clear that Maxwell's theory is not the only one to describe the electromagnetic field. One of the most successful generalizations is the generalized electrodynamics \cite{ref15}. Actually, Podolsky's theory is the only one linear, and Lorentz and $U(1)$ invariant generalization of Maxwell's theory \cite{ref17}. Another interesting feature inherent to Podolsky's theory is the existence of a generalized gauge condition, namely, the generalized Lorenz condition: $\Omega[A]=\left(1+M^{-2}\square\right)\partial_\mu A^{\mu}$; it must be considered an important issue, since it is only through the choice of the correct gauge condition that we can completely fix the gauge degrees of freedom of a given gauge theory \cite{ref16}.

On the other hand, the generalized electrodynamics when interacting with matter fields has been studied in details at finite \cite{ref18} and zero temperatures \cite{ref19}, and also has revealed to be a compelling theory, leading to interesting results at radiative corrections \cite{ref19}. Therefore, based on the recent results at finite temperature literature, we believe it to be a rather natural and stimulating extension to study the thermal generalized electrodynamics in a static gravitation background. The main purpose of the present work is to obtain the higher-loop corrections to the effective Lagrangian of a generalized electrodynamics plasma in a static gravitation background. It should be emphasized that the background metric is \textit{static} when it does not depend on time. The one-loop results do not take into account the interactions between electrons and (generalized) photons. In order to consider these effects we shall compute the two-loop contribution.

In this paper, we discuss a generalized electrodynamics in a static gravitation background at thermodynamical equilibrium in the light of the Matsubara imaginary-time formalism \cite{ref21,ref22}. In Sect.\ref{sec:1} we start by making a brief review of the generalized electrodynamics and presenting some useful results from the vierbein formalism \cite{ref23}. Moreover, we evaluate in detail the three contributions from the one-particle irreducible diagrams at one-loop order for
the effective Lagrangian. Next, in Sect.\ref{sec:2}, we discuss and present a detailed calculation of the two-loop effective Lagrangian, where some approximation are needed in order to perform the integration exactly. Besides, we discuss some general remarks about the infrared behavior of the present theory, as well as some nonperturbative effects. In Sect.\ref{sec:3} we summarize the results, and present our final remarks and prospects.

\section{Effective Lagrangian at one-loop order}

\label{sec:1}

In this section we will introduce our basic notation and describe the analysis method. Let us consider the Lagrangian density for (generalized) photons and electrons in a gravitational background \cite{ref19}
\begin{align}
\mathcal{L} =&\sqrt{\left\vert g\right\vert }\bigg\{i\bar{\psi}g^{\mu \nu
}\gamma _{\mu }\left( \partial _{\nu }-ieA_{\nu }\right) \psi +g^{\mu \nu }\partial _{\mu }\bar{c}\left(
1+M^{-2}\square \right) \partial _{\nu }c\notag \\
&-\frac{1}{4}g^{\mu \nu }g^{\alpha \beta }F_{\mu \alpha }F_{\nu \beta }+\frac{1}{2M^{2}}%
g^{\mu \sigma }g^{\alpha \lambda }g^{\nu \xi }\partial _{\mu }F_{\sigma
\lambda }\partial _{\xi }F_{\nu \alpha }  -\frac{1}{2\xi }\left( \left( 1+M^{-2}\square \right) g^{\mu \nu }\partial
_{\mu }A_{\nu }\right) ^{2}\bigg\},\label{eq 0.0}
\end{align}%
where $g_{\mu \nu }$ is the metric tensor $\left( \left\vert g\right\vert=\left\vert \det g_{\mu \nu }\right\vert \right) $
and $M$ is the free parameter from the generalized electrodynamics, also $F_{\mu \nu }=\partial_{\mu }A_{\nu }-\partial _{\nu }A_{\mu }$ is the electromagnetic stress-tensor, and $\left( c,\bar{c}\right) $ is the set of the Faddeev-Popov ghost fields. It should be remarked that in order to determine a closed form expression for the static effective Lagrangian at the high-temperature limit we shall consider some approximations. For this purpose, we can bear in mind the much larger scale of temperature and thus neglect the suppressed quantities such as all the spacetime derivatives of the metric as well as the fermion mass.

\begin{figure*}[tbp]
\begin{center}
\includegraphics[scale=0.9]{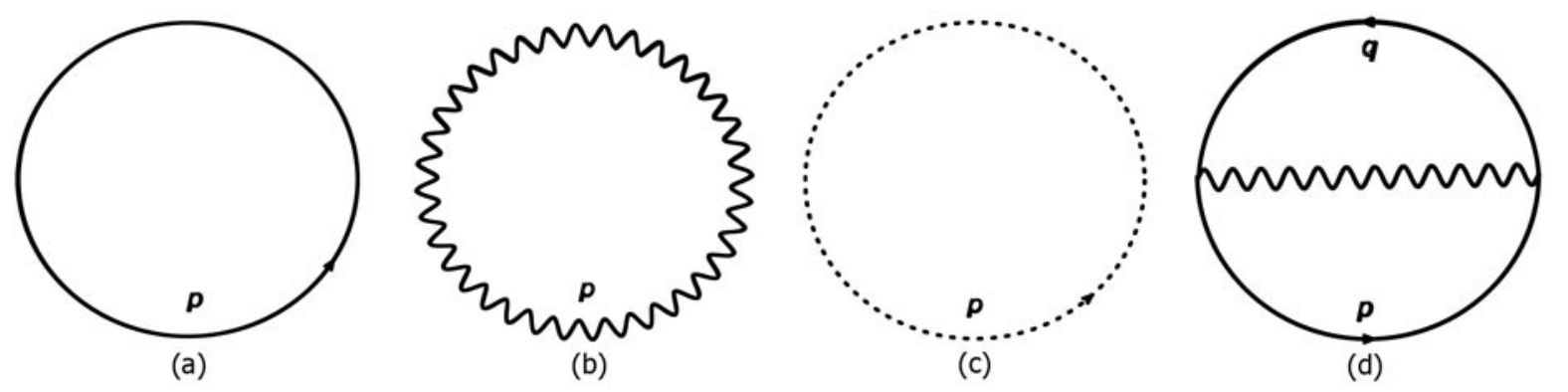}
\end{center}

\caption{One-loop and two-loop diagrams contribution to the effective Lagrangian.}
\label{fig1}
\end{figure*}

To derive the Feynman rules for the dynamical fields in a gravitational background is suitable to
make use of the vierbein formalism \cite{ref23}. In fact, a local Lorentz frame can be defined in terms of the
vierbein $e_{\mu }^{~~a}$, so that in a given point of the manifold the metric can be written as%
\begin{equation}
g_{\mu \nu }=e_{\mu }^{~~a}e_{\nu }^{~~b}\eta _{ab},  \label{eq 0.1}
\end{equation}%
where the greek and latin indices stand for general and local coordinates, respectively. Moreover, we shall make use throughout the paper of the notation $\tilde{p}_{a}=e^{~~\mu }_{a}p_{\mu }$. \footnote{From this very definition it follows: $ \square =g^{\mu \nu }\partial _{\mu}\partial _{\nu }=\eta ^{ab}\tilde{\partial}_{a}\tilde{\partial}_{b}=\widetilde{\square }$.} Hence, this formalism allows us to write the Lagrangian density \eqref{eq 0.0} in the following form%
\begin{align}
\mathcal{L} =&\sqrt{\left\vert g\right\vert }\bigg\{i\bar{\psi}\tilde{\gamma}^{a}\left( \tilde{\partial}_{a}
-ig\tilde{A}_{a}\right) \psi -\frac{1}{4}\tilde{F}_{ab}\tilde{F}^{ab}\notag\\
&+\frac{1}{2M^{2}}\tilde{\partial}_{a}\tilde{F}%
^{ab}\tilde{\partial}^{c}\tilde{F}_{cb} -\frac{1}{2\xi }\left( \left( 1+M^{-2}\widetilde{\square }\right) \tilde{\partial}_{a}\tilde{A}^{a}\right) ^{2} +\tilde{\partial}_{a}\bar{c}\left(1+M^{-2}\widetilde{\square }\right) \tilde{\partial}^{a}c\bigg\}, \label{eq 0.2}
\end{align}%
in such vierbein basis we have that the Dirac matrices $\tilde{\gamma}_{a}$ satisfy%
\begin{equation}
\left\{ \tilde{\gamma}_{a},\tilde{\gamma}_{b}\right\} =2\eta _{ab}=2g_{\mu
\nu }e_{a}^{~~\mu }e_{b}^{~~ \nu }.  \label{eq 0.3}
\end{equation}%
By means of completeness, we shall evaluate the lowest order contributions to the effective Lagrangian. The respective contributions are presented at Fig.\ref{fig1}. These are the diagrams from the fermion loop $(a)$, generalized photon loop $(b)$ and ghost loop $(c)$%
\begin{equation}
\mathcal{L}_{1}=\mathcal{L}_{1}^{F}+\mathcal{L}_{1}^{P}+\mathcal{L}_{1}^{G},
\label{eq 0.4}
\end{equation}%
with%
\begin{align}
\mathcal{L}_{1}^{F} =&\ln {\det}_{D}\left( i\tilde{\gamma}^{a}\tilde{\partial}%
_{a}\right) ,  \label{eq 0.5} \\
\mathcal{L}_{1}^{P} =&-\frac{1}{2}\ln {\det}_{L} \left( M_{ab}\right) ,
\label{eq 0.6} \\
\mathcal{L}_{1}^{G} =&\ln \det \left( \left( 1+M^{-2}\widetilde{\square }%
\right) \widetilde{\square }\right) ,  \label{eq 0.7}
\end{align}%
where%
\begin{align}
M_{ab}\left( x,y\right) =\left[ \eta _{ab}\widetilde{\square }-\left\{ 1-%
\frac{1}{\xi }\left( 1+M^{-2}\widetilde{\square }\right) \right\} \tilde{%
\partial}_{a}\tilde{\partial}_{b}\right] \times\left( 1+M^{-2}\widetilde{\square }%
\right) \delta \left( x,y\right) .  \label{eq 0.8}
\end{align}%
We denoted $\det_{D}$ as being the determinant over the Dirac matrices and the Hilbert space ($\det $), but notice that in $\mathcal{L}_{1}^{P}$ we also have the determinant on the spacetime indices. Now, we can solve it using the general result $\det_{L}\left(M_{ab}\right) =\det \left\{\left[ \left( 1+M^{-2}\widetilde{\square }\right) \widetilde{\square }\right] ^{\omega }\left( 1+M^{-2}\widetilde{\square }\right)\right\} $, leading to%
\begin{align}
\mathcal{L}_{1}^{F} =&\ln {\det}_{D}\left( i\tilde{\gamma}^{a}\tilde{\partial}%
_{a}\right) , \\
\mathcal{L}_{1}^{P} =&-\frac{\omega }{2}\ln \det \left( \left( 1+M^{-2}%
\widetilde{\square }\right) \widetilde{\square }\right) -\frac{1}{2}\ln \det \left( 1+M^{-2}\widetilde{\square }\right) , \\
\mathcal{L}_{1}^{G} =&\ln \det \left( \left( 1+M^{-2}\widetilde{\square }%
\right) \widetilde{\square }\right) .
\end{align}%
Moreover, using the imaginary-time formalism \cite{ref21,ref22} we can express these three contributions as%
\begin{align}
\mathcal{L}_{1}^{F} =&\frac{1}{2\beta }2^{E\left( \omega/2\right) }\underset{n_{F}%
}{\sumint  }\ln \left( -\beta ^{2}\eta ^{ab}\tilde{p}_{a}\tilde{p}_{b}\right) ,\label{eq 0.9} \\
\mathcal{L}_{1}^{P}+\mathcal{L}_{1}^{G} =&\frac{1-\omega }{2\beta }\underset{%
n_{B}}{\sumint  }\ln \left[ \beta ^{2}\left( 1-\frac{\eta ^{ab}\tilde{p}_{a}\tilde{p}_{b}}{M^{2}}%
\right) \right]  +\frac{2-\omega }{2\beta }\underset{n_{B}}{\sumint }\ln \left[ -\beta ^{2}\eta ^{cd}\tilde{p}%
_{c}\tilde{p}_{d}\right] , \label{eq 0.11}
\end{align}
where we have defined the notation for the fermionic and bosonic sum/integral
\begin{equation}
\underset{n_{F}}{\sumint  } \equiv \underset{n_{F}}{\sum  }\int \frac{d^{\omega -1}p}{\left( 2\pi \right) ^{\omega -1}}, \quad
\underset{n_{B}}{\sumint  } \equiv \underset{n_{B}}{\sum  } \int \frac{d^{\omega -1}p}{\left( 2\pi \right) ^{\omega -1}}.
\end{equation}
It should be emphasized that we are employing the \emph{irreducible} representation for the Dirac matrices, so that the trace of the identity is given by $tr\left( I\right) =2^{E\left(\omega /2\right) }$, where $E\left( \omega /2\right) $ is the integer part of $\omega /2$. Furthermore, notice that in the first expression, Eq.\eqref{eq 0.9}, the sum is over the fermionic Matsubara frequencies $\omega _{n_{F}}=\frac{\left( 2n_{F}+1\right) \pi }{\beta }$, where the time component of the momentum is $p_{0}=i\omega _{n_{F}}$, while, in the second expression, Eq.\eqref{eq 0.11}, the sum is over bosonic Matsubara frequency $\omega _{n_{B}}=\frac{2n_{B}\pi }{\beta }$ $\left(p_{0}=i\omega _{n_{B}}\right)$.

In order to evaluate the sum/integrals in Eqs.\eqref{eq 0.9} and \eqref{eq
0.11} in a simpler way, one can conveniently choose a \textit{locally rest vierbein frame} as defined \cite{ref13,ref23}%
\begin{equation}
\tilde{p}_{0}=\frac{p_{0}}{\sqrt{g_{00}}}.  \label{eq 0.12}
\end{equation}%
At finite temperature, the thermal bath introduces a privileged reference frame which may be characterized by
its four-velocity $u^{\mu }$ \cite{ref21,ref22}. In all points of the manifold, we have a special coordinate system (locally rest
frame) in which%
\begin{equation}
u^{\mu }=\left( \frac{1}{\sqrt{g_{00}}},\overrightarrow{0}\right) .
\end{equation}%
Proceeding this way, we may use%
\begin{equation}
\tilde{p}_{a}\tilde{u}^{a}=p_{\mu }u^{\mu }=\frac{p_{0}}{\sqrt{g_{00}}},
\end{equation}%
to define a special class of vierbein, as shown in \eqref{eq 0.12}. Hence,
for two arbitrary four-vectors, we have that their scalar product is given by \cite{ref13}%
\begin{align}
\eta ^{ab}\tilde{p}_{a}\tilde{q}_{b}=&g^{\mu \nu }p_{\mu }q_{\nu },\notag \\
=&\frac{p_{0}q_{0}}{g_{00}}+g^{ij}\left[ p_{i}+
\left( g^{-1}\right) _{im}g^{0m}p_{0}\right] \left[ q_{j}+\left( g^{-1}\right) _{jl}g^{0l}q_{0}\right] . \label{eq 0.13}
\end{align}%
In particular, from these results it follows that
\begin{equation}
\eta ^{ab}\tilde{p}_{a}\tilde{p}_{b}=g^{\mu \nu }p_{\mu }p_{\nu }=\left(\frac{1}{\sqrt{g_{00}}}\right) ^{2}
\left[ \left( p_{0}\right)^{2}-\left\vert p^{\prime }\right\vert ^{2}\right] ,  \label{eq 0.15}
\end{equation}%
where we have introduced the change of variables
\begin{equation}
p_{i}^{\prime }=\sqrt{g_{00}}\left[ N_{i}^{~~j}p_{j}+\left(g^{-1}\right) _{im}g^{0m}p_{0}\right] , \label{eq 0.16}
\end{equation}%
in such a way that $g^{ij}=N_{~~ l}^{i}\eta ^{lm}N_{m}^{~~ j}$. Moreover, under this change of variables we have the nontrivial Jacobian: $d^{\omega -1}p^{\prime }= \mathcal{J}d^{\omega -1}p$, with $\mathcal{J}=\left( g_{00}\right) ^{\frac{\omega-1}{2}}\sqrt{\bar{g}}=\frac{\left( g_{00}\right)
^{\frac{\omega }{2}}}{\sqrt{\left\vert g\right\vert }}$, where we have made use of the identity $\bar{g}^{-1}g_{00}=g$, in which $\bar{g}= \det g^{ij}$ \cite{ref13}. Therefore, from the result \eqref{eq 0.15} and subsequent manipulation on the change of variables \eqref{eq 0.16}, we can deal properly with the metric dependence, and the expressions \eqref{eq 0.9} and \eqref{eq 0.11} are written simply as
\begin{align}
\mathcal{L}_{1}^{F}= \frac{2^{E\left( \omega /2\right) }}{2\beta }\frac{\sqrt{\left\vert g\right\vert }}{\left( g_{00}\right)
^{\frac{\omega }{2}}}\underset{n_{F}}{\sumint }\ln \left( -\beta ^{2}\left[ \left( q_{0}\right)^{2}-\left\vert q\right\vert ^{2}\right] \right) ,  \label{eq 0.18}
\end{align}
and
\begin{align}
\mathcal{L}_{1}^{P}+\mathcal{L}_{1}^{G}&= \frac{1-\omega }{2\beta }\frac{\sqrt{\left\vert g\right\vert }}{\left( g_{00}\right)
^{\frac{\omega }{2}}}\underset{n_{B}}{\sumint }\ln \left[ \beta ^{2}\left( 1-\frac{\left( q_{0}\right) ^{2}-\left\vert q\right\vert ^{2} }{\left(\sqrt{g_{00}} M\right)^{2}}\right) \right]\notag\\
&+\frac{2-\omega }{2\beta }\frac{\sqrt{\left\vert g\right\vert }}{\left(g_{00}\right) ^{\frac{\omega }{2}}}\underset{n_{B}}
{\sumint }\ln \left[ -\beta ^{2}\left[
\left( q_{0}\right) ^{2}-\left\vert q\right\vert ^{2}\right] \right] . \label{eq 0.19}
\end{align}%
We should emphasize that all temperature-independent parts of \eqref{eq 0.18} and \eqref{eq 0.19} lead to a divergent result, i.e., the zero-point energy of the vacuum, which can be subtracted off since it is an unobservable constant.

We shall start by evaluating first the fermionic sum and integral from \eqref{eq 0.18} by means of imaginary-time formalism \cite{ref21,ref22}
\begin{align}
I_{F_1}^{\left( 1\right) }=\underset{n_{F}}{\sumint }\ln \left( -\beta ^{2}\left[ \left(
q_{0}\right) ^{2}-\left\vert q\right\vert ^{2}\right] \right)=2\int \frac{d^{\omega -1}q}{\left( 2\pi \right) ^{\omega -1}}\ln \left( 1+e^{-\beta
\omega _{q}}\right) ,
\end{align}%
moreover, the remaining integral can be solved by using standard rules of finite temperature integration. For instance, we may use the well-known result for fermionic fields%
\begin{equation}
\int_{0}^{\infty }\frac{z^{x-1}}{1+e^{z}}dz=\left( 1-2^{1-x}\right) \Gamma
\left( x\right) \zeta \left( x\right) ,  \label{eq 0.20}
\end{equation}%
and then obtain
\begin{equation}
I_{F_1}^{\left( 1\right) }=\frac{2\beta ^{1-\omega }}{\left( 4\pi \right) ^{%
\frac{\omega -1}{2}}}\frac{\Gamma \left( \omega \right) \zeta \left( \omega
\right) }{\Gamma \left( \frac{\omega +1}{2}\right) }\left( 1-2^{1-\omega
}\right) .  \label{eq 0.21}
\end{equation}%
Hence, from the result \eqref{eq 0.21} we have the following expression for the fermionic contribution \eqref{eq 0.18}
\begin{equation}
\mathcal{L}_{1}^{F}=\sqrt{\left\vert g\right\vert }\left( \frac{1}{\beta
\sqrt{g_{00}}}\right) ^{\omega }\left[ \frac{2^{E\left( \omega /2\right)
}\left( 1-2^{1-\omega }\right) }{\left( 2\sqrt{\pi }\right) ^{\omega -1}}%
\frac{\Gamma \left( \omega \right) \zeta \left( \omega \right) }{\Gamma
\left( \frac{\omega +1}{2}\right) }\right] .  \label{eq 0.22}
\end{equation}%
Nevertheless, the massless bosonic sum/integral from \eqref{eq 0.19} can be evaluate by similar computation
\begin{align}
I_{B_{1}}^{\left( 1\right) }=\underset{n_{B}}{\sumint }
\ln \left( -\beta ^{2}\left[ \left(q_{0}\right) ^{2}-\left\vert q\right\vert ^{2}\right] \right)=-\frac{2\beta ^{1-\omega }}{\left( 4\pi \right) ^{\frac{\omega -1}{2}}}\frac{\Gamma \left(
\omega \right) \zeta \left( \omega \right) }{\Gamma \left( \frac{\omega +1}{2}\right) },  \label{eq 0.23}
\end{align}%
in which we have made use of the well-known result%
\begin{equation}
\int_{0}^{\infty }\frac{z^{x-1}}{e^{z}-1}dz=\Gamma \left( x\right) \zeta
\left( x\right) .
\end{equation}%
At last, we shall evaluate the massive bosonic sum/integral \cite{ref18} \footnote{%
Since $\beta \sqrt{q^{2}+M^{2}}>0$, thus $e^{-\beta \sqrt{q^{2}+M^{2}}}<1$,
it then follows that $\frac{e^{-\beta \sqrt{q^{2}+M^{2}}}}{1-e^{-\beta \sqrt{%
q^{2}+M^{2}}}}=\underset{k=1}{\sum }e^{-k\beta \sqrt{q^{2}+M^{2}}}.$}%
\begin{align}
I_{B_{2}}^{\left( 1\right) }=&\underset{n_{B}}{\sumint }
\ln \left[ \beta ^{2}\left( 1-\frac{\left( q_{0}\right) ^{2}-\left\vert q\right\vert ^{2} }{\left(\sqrt{g_{00}} M\right)^{2}}\right) \right], \notag \\
=&-\frac{2\beta }{\left( 4\pi \right) ^{\frac{\omega -1}{2}}}\frac{\left(\sqrt{g_{00}} M\right)^{\omega }}{\Gamma \left( \frac{\omega +1}{2}\right) } \underset{k=1}{\sum }\int_{1}^{\infty }dw\left( w^{2}-1\right) ^{\frac{\omega -1}{2}}e^{-k\beta \sqrt{g_{00}} M w}.
\end{align}%
Moreover, one may use the integral representation for the modified Bessel function of the second-kind \cite{ref24} to solve the above integral, and obtain
\begin{equation}
\frac{\sqrt{\pi }}{\Gamma \left( \frac{\omega +1}{2}\right) }\int_{1}^{\infty }dx\left( x^{2}-1\right) ^{\frac{\omega -1}{2}}e^{-A x}
=\left( \frac{2}{A}\right) ^{\frac{\omega }{2}}K_{\frac{\omega }{2}}\left( A\right) .
\end{equation}%
Finally, it follows that we can write the massive bosonic determinant in terms of a perturbative series
\begin{equation}
I_{B_{2}}^{\left( 1\right) }=-\frac{4\beta \left(\sqrt{g_{00}} M\right)^{\omega }}{\left( 2\pi \right)^{\frac{\omega }{2}}}\underset{k=1}{\sum }
\left( \frac{1}{k\beta \sqrt{g_{00}} M}\right) ^{\frac{\omega }{2}}K_{\frac{\omega }{2}}\left( k\beta \sqrt{g_{00}} M\right) .\label{eq 0.24}
\end{equation}%
Hence, collecting the results \eqref{eq 0.23} and \eqref{eq 0.24} the bosonic contribution \eqref{eq 0.19} reads
\begin{align}
\mathcal{L}_{1}^{P}+\mathcal{L}_{1}^{G} =&\left( \omega -1\right) \frac{\sqrt{\left\vert g\right\vert }}
{\left( g_{00}\right) ^{\frac{\omega }{2}}}\frac{2}{\left( 2\pi \right) ^{\frac{\omega }{2}}}
\underset{k=1}{\sum }\left( \frac{\sqrt{g_{00}} M}{k\beta }\right) ^{\frac{\omega }{2}}K_{\frac{\omega }{2}}\left( k\beta \sqrt{g_{00}} M\right)  \notag \\
&+\left( \omega -2\right) \frac{\sqrt{\left\vert g\right\vert }}{\left( 2\sqrt{\pi }\right)^{\omega -1}}
\left( \frac{1}{\beta \sqrt{g_{00}}}\right) ^{\omega }\frac{\Gamma \left( \omega \right) \zeta \left( \omega \right)
}{\Gamma \left( \frac{\omega +1}{2}\right) }.  \label{eq 0.25}
\end{align}%
Therefore, with the results, Eqs.\eqref{eq 0.22} and \eqref{eq 0.25}, one
may then write the complete one-loop effective Lagrangian \eqref{eq 0.4} as
\begin{align}
\mathcal{L}_{1} &=\sqrt{\left\vert g\right\vert }\left( \frac{1}{\beta \sqrt{g_{00}}}\right) ^{\omega }
\left[ \frac{1}{\left( 2\sqrt{\pi }\right)^{\omega -1}}\frac{\Gamma \left( \omega \right) \zeta \left( \omega \right)
}{\Gamma \left( \frac{\omega +1}{2}\right) }\right] \notag \\
&\times \bigg\{2^{E\left( \omega
/2\right) }\left( 1-2^{1-\omega }\right) +\left( \omega -2\right) +\frac{1}{\sqrt{\pi }}\frac{\Gamma \left( \frac{\omega +1}{2}\right) }{\Gamma \left( \omega -1\right) \zeta \left(
\omega \right) }\underset{k=1}{\sum }\left( \frac{2\beta \sqrt{g_{00}} M}{k}\right) ^{\frac{\omega }{2}}K_{\frac{\omega }{%
2}}\left( k\beta \sqrt{g_{00}} M\right) \bigg\}.  \label{eq 0.26}
\end{align}%
Unfortunately, there is not known a closed form for the above series, which means that the thermal properties may be used in such a way to find a suitable approximation in order to evaluate its expression. For this matter, we may assume the situation where $\beta M\gg 1$, which means that the parameter $M$ should be much larger than the thermal energy. In this case we may use the asymptotic expansion for $\left\vert z\right\vert \rightarrow \infty $ \cite{ref24}
\begin{equation}
K_{\nu }\left( z\right) \sim \sqrt{\frac{\pi }{2z}}e^{-z}\left[ 1+\frac{4\nu
^{2}-1}{8z}+....\right] .
\end{equation}%
This results into
\begin{equation}
\underset{k=1}{\sum }\left( \frac{2\beta \sqrt{g_{00}} M}{k}\right) ^{\frac{\omega }{2}}K_{%
\frac{\omega }{2}}\left( k\beta \sqrt{g_{00}} M\right) \simeq \underset{k=1}{\sum }\left(
\frac{2\beta \sqrt{g_{00}} M}{k}\right) ^{\frac{\omega }{2}}\sqrt{\frac{\pi }{2k\beta \sqrt{g_{00}} M}}e^{-k\beta \sqrt{g_{00}} M}.
\end{equation}%
Moreover, as aforementioned, we have that $\left( e^{-\beta M}\right) ^{n+1}\ll \left(e^{-\beta M}\right) ^{n}$. Hence, it follows that the leading term of the above sum is for $k=1$,%
\begin{equation}
\underset{k=1}{\sum }\left( \frac{2\beta \sqrt{g_{00}} M}{k}\right) ^{\frac{\omega }{2}}K_{%
\frac{\omega }{2}}\left( k\beta \sqrt{g_{00}} M\right) 
\simeq \sqrt{\pi}\left( 2\beta \sqrt{g_{00}} M\right) ^{\frac{\omega -1}{2}}e^{-\beta \sqrt{g_{00}} M}.  \label{eq 0.27}
\end{equation}%
Therefore, within this approximation, we have that the Eq.\eqref{eq 0.26} reads%
\begin{align}
\mathcal{L}_{1} =&\sqrt{\left\vert g\right\vert }\left( \frac{1}{\beta \sqrt{%
g_{00}}}\right) ^{\omega }\left[ \frac{1}{\left( 2\sqrt{\pi }\right)
^{\omega -1}}\frac{\Gamma \left( \omega \right) \zeta \left( \omega \right)
}{\Gamma \left( \frac{\omega +1}{2}\right) }\right]  \notag \\
&\times \bigg\{2^{E\left( \omega /2\right) }\left( 1-2^{1-\omega }\right)
+\left( \omega -2\right) +\frac{\Gamma \left( \frac{\omega +1}{2}\right) }{%
\Gamma \left( \omega -1\right) \zeta \left( \omega \right) }\left( 2\beta
\sqrt{g_{00}} M\right) ^{\frac{\omega -1}{2}}e^{-\beta \sqrt{g_{00}} M}\bigg\}.  \label{eq 0.28}
\end{align}%
As one should expected for a density, either Eq.\eqref{eq 0.26} or Eq.\eqref{eq 0.28} presents the factor $\sqrt{\left\vert g\right\vert }$. They also exhibit a characteristic temperature dependence in the full expression on the \textit{local temperature} defined by $T_{loc}=\frac{T}{\sqrt{g_{00}}}$, which is a direct consequence of the Tolman-Ehrenfest effect (thermal time) \cite{ref14}. On the other hand, this result is surprisingly interesting and unexpected, due to the fact that we have added a scale by the massive contribution. Therefore, these facts yield to the resulting expression to be invariant under the scale transformation $g_{\mu \nu}\rightarrow \Omega g_{\mu \nu }$ \cite{ref25}.


\section{Two-loop Effective Lagrangian}

\label{sec:2}

In order to investigate the effect of higher-order correction, we shall evaluate the two-loop order contribution to the effective Lagrangian as depicted in diagram $(d)$ at the Fig.\ref{fig1}. We will apply the same technique as illustrated before in the calculation of the one-loop contribution. This can be obtained by computing the contribution
\begin{align}
\mathcal{L}_{2}=\frac{e^{2}}{2\beta ^{2}}\frac{1}{\sqrt{\left\vert g\right\vert }}\underset{n_{F}}{\sumint }\underset{m_{F}}{\sumint }
 tr\left[ \gamma ^{\mu }\frac{1}{ \gamma .p} \gamma ^{\nu}\frac{1}{ \gamma .q} \right]\left[ g _{\mu \nu }-\frac{k_{\mu }k_{\nu }}{k^{2}}\right] \left[ \frac{1}{k^{2}}-\frac{1}{k^{2}-M^{2}}\right] ,  \label{eq 2.1}
\end{align}%
where $k=p-q$ is the photon momentum; besides, we are using the gauge field propagator expression at Landau gauge $\xi =0$ \cite{ref19},%
\begin{equation}
iD_{\mu \nu }=\left[ g _{\mu \nu }-\frac{k_{\mu }k_{\nu }}{k^{2}}\right] %
\left[ \frac{1}{k^{2}}-\frac{1}{k^{2}-M^{2}}\right] .
\end{equation}%
Moreover, in the vierbein basis, we have $\left\{ \tilde{\gamma}_{a},\tilde{\gamma}%
_{b}\right\} =2\eta _{ab}$, then the Dirac $\tilde{\gamma}$ matrices trace may readily be evaluated%
\begin{equation}
tr\left[ \tilde{\gamma}^{a}\tilde{\gamma}^{b}\tilde{\gamma}^{c}%
\tilde{\gamma}^{d}\right] =2^{E\left( d/2\right) }\left( \eta ^{ab}\eta
^{cd}-\eta ^{ac}\eta ^{bd}+\eta ^{ad}\eta ^{bc}\right) ,  \label{eq 2.2}
\end{equation}%
and, after some algebraic manipulation, we obtain the resulting expression for $\mathcal{L}_{2}$ from \eqref{eq 2.1}
\begin{align}
\mathcal{L}_{2}= \frac{e^{2}}{\beta ^{2}}\frac{2^{E\left( d/2\right) -1}}{\sqrt{\left\vert g\right\vert }}
\underset{n_{F}}{\sumint } \underset{m_{F}}{\sumint }  \left[ \frac{1}{\tilde{k}^{2}}-\frac{1}{\tilde{k}^{2}-M^{2}}\right] \left[ \left( 3-\omega \right) \frac{\left( \tilde{p}.\tilde{q}\right) }{2\tilde{q}^{2}\tilde{p}^{2}}
-\frac{2\left( \tilde{p}.\tilde{q}\right) }{\tilde{k}^{2}\tilde{p}^{2}}+\frac{1}{\tilde{k}^{2}}+\frac{\left(
\tilde{p}.\tilde{q}\right) ^{2}}{\tilde{k}^{2}\tilde{q}^{2}\tilde{p}^{2}}%
\right] .\label{eq 2.4}
\end{align}%
However, at finite temperature, the massive term is not easily handled, neither in order to get a closed expression for it. Hence, as in the Sect.\ref{sec:1}, we shall regard henceforth the approximation $M^2/k^2 \approx \beta ^2 M^2 \gg 1$ in \eqref{eq 2.4}, which is consistent with the hard thermal loop approximation, and then consider the leading terms
\begin{equation}
\frac{1}{\tilde{k}^{2}}-\frac{1}{\tilde{k}^{2}-M^{2}}=\frac{1}{\tilde{k}^2}\frac{1}{1-\tilde{k}^2/M^2}\approx \frac{1}{\tilde{k}^{2}}+\frac{1}{M^{2}}+\frac{\tilde{k}^{2}}{M^{4}},
\end{equation}
in order to get the most significant contribution from the generalized electrodynamics. Thus, this yields to the following expression%
\begin{align}
\mathcal{L}_{2}= \frac{e^{2}}{\beta ^{2}}\frac{2^{E\left( d/2\right) -1}%
}{\sqrt{\left\vert g\right\vert }}\underset{n_{F}}{\sumint }\underset{m_{F}}{\sumint } \left[\frac{1}{\tilde{k}^{2}}+\frac{1}{M^{2}}+\frac{\tilde{k}^{2}}{M^{4}}\right]\left[ \left( 3-\omega \right) \frac{\left( \tilde{p}.\tilde{q}%
\right) }{\tilde{q}^{2}\tilde{p}^{2}}-\frac{4\left( \tilde{p}.\tilde{q}\right) }
{\tilde{k}^{2}\tilde{p}^{2}}+\frac{2}{\tilde{k}^{2}}+\frac{2\left( \tilde{p}.%
\tilde{q}\right) ^{2}}{\tilde{k}^{2}\tilde{q}^{2}\tilde{p}^{2}}\right] .  \label{eq 2.5}
\end{align}%
In particular, as matter of illustration, an useful manipulation in order for future calculation reads
\begin{equation}
\underset{n_{F}}{\sumint }\underset{m_{F}}{\sumint }\frac{2\left( \tilde{p}.\tilde{q}\right) }{\tilde{q}^{2}\tilde{p}^{2}\tilde{k}^{2}}%
=2I_{F}^{\left( 1\right) }I_{B}^{\left( 1\right) }-\left( I_{F}^{\left(1\right) }\right) ^{2},  \label{eq 2.6}
\end{equation}%
where we will use the following definition henceforth for the fermionic and bosonic quantities
\begin{align}
I_{F}^{\left( n\right) }= \underset{n_{F}}{\sumint }\frac{1}{\left( \tilde{p}^{2}\right) ^{n}}, \quad I_{B}^{\left( n\right) }= \underset{l_{B}}{\sumint }\frac{1}{\left( \tilde{k}^{2}\right) ^{n}}. \label{eq 2.6b}
\end{align}%
We can proceed in the exact same way as outlined above, using results from dimensional regularization technique, and then rewrite conveniently all the terms from Eq.\eqref{eq 2.5} in its dominant contribution as the following
\begin{align}
\mathcal{L}_{2}=e^{2}\frac{2^{E\left( d/2\right) -1}}{\sqrt{\left\vert
g\right\vert }}\left( \omega -2\right) \bigg\{- I_{F}^{\left( 1\right) }I_{B}^{\left( 1\right) }+\frac{1}{2}\left(I_{F}^{\left( 1\right) }\right)^2+\frac{1}{2M^4} \left[ \frac{1}{\beta ^{2}}\underset{n_{F}}{\sumint }\underset{m_{F}}{\sumint } \frac{4\left( \tilde{p}.\tilde{q}\right) ^{2}}{\tilde{p}^{2}\tilde{q}^{2}}\right] \bigg\}.
\label{eq 2.12}
\end{align}%
The evaluation of the sum and integral from the last term of Eq.\eqref{eq 2.12} is lengthy but straightforward, and it follows directly from the aforementioned identities and well-known results at finite temperature \cite{ref21,ref22}, and the result reads
\begin{align}
I_{F}  =&\frac{1}{\beta^{2}}\underset{n_{F},m_{F}}{\sum}\int\frac{d^{\omega-1}p}{\left(2\pi\right)^{\omega-1}}\frac{d^{\omega-1}q}{\left(2\pi\right)^{\omega-1}}\frac{4 \left(\tilde{p}.\tilde{q}\right)^{2}}{\tilde{p}^{2}\tilde{q}^{2}},\nonumber \\
  =&\frac{32}{3\sqrt{\pi}}\frac{\left(\sqrt{\left\vert g\right\vert }\right)^{2}}{\left(\sqrt{g_{00}}\beta \right)^{2\omega}}\frac{1}{\left(2\sqrt{\pi}\right)^{2\left(\omega-1\right)}}\frac{\left[\left(1-2^{1-\omega}\right)\Gamma\left(\omega\right)
 \zeta\left(\omega\right)\right]^{2}}{\Gamma\left(\frac{\omega-1}{2}\right)\Gamma\left(\frac{\omega-2}{2}\right)},
\end{align}
the remaining terms may also be easily evaluated,
\begin{equation}
I_{F}^{\left(1\right)}=\frac{2\sqrt{\left\vert g\right\vert }}{\left(\beta\sqrt{g_{00}}\right)^{\omega-2}}\frac{\left(1-2^{3-\omega}\right)\Gamma\left(\omega-2\right)
\zeta\left(\omega-2\right)}{\Gamma\left(\frac{\omega-1}{2}\right)\left(2\sqrt{\pi}\right)^{\omega-1}},
\end{equation}
and
\begin{equation}
I_{B}^{\left(1\right)}=-\frac{2\sqrt{\left\vert g\right\vert }}{\left(\beta\sqrt{g_{00}}\right)^{\omega-2}}\frac{\Gamma\left(\omega-2\right)\zeta\left(\omega-2\right)}{\Gamma\left(\frac{\omega-1}{2}\right)\left(2\sqrt{\pi}\right)^{\omega-1}},
\end{equation}
where we have dropped off the temperature-independent terms. Therefore, after evaluating all the terms, we obtain the following expression for the two-loop effective Lagrangian
\begin{align}
\mathcal{L}_{2} & =\frac{2^{E\left(d/2\right)}e^{2}\sqrt{\left\vert g\right\vert }}{\left(\beta\sqrt{g_{00}}\right)^{2\omega-4}}\frac{\left(\omega-2\right)}{\left[\Gamma\left(\frac{\omega-1}{2}\right)\left(2\sqrt{\pi}\right)^{\left(\omega-1\right)}\right]^2} \notag \\
&\times \biggl\{\left(1-2^{3-\omega}\right)
\left(3-2^{3-\omega}\right)\Gamma^{2}\left(\omega-2\right)\zeta^{2}\left(\omega-2\right)\notag\\
& +\frac{8}{3\sqrt{\pi}}\frac{1}{\left(\beta\sqrt{g_{00}}M\right)^{4}}
 \frac{\Gamma\left(\frac{\omega-1}{2}\right)}{\Gamma\left(\frac{\omega-2}{2}\right)}\left[\left(1-2^{1-\omega}\right)\Gamma\left(\omega\right)\zeta\left(\omega\right)\right]^{2}\biggr\}, \label{eq 2.28}
\end{align}
The result given by Eq.\eqref{eq 2.28} can be identified with the two-loop Podolsky contribution to the pressure, in the high-temperature regime; besides, it presents, as expected for a density, the factor $\sqrt{\left\vert g\right\vert }$. Furthermore, as the one-loop result, the two-loop expression also displays a simple dependence on the Tolman local temperature, as defined in $T_{loc}=\frac{T}{\sqrt{g_{00}}}$. Moreover, from Eq.\eqref{eq 2.28} we may identify the first term as being the two-loop QED contribution as calculated in \cite{ref13}
\begin{align}
\mathcal{L}_{2}&=\sqrt{\left\vert g\right\vert }\left[ \frac{e^{2}}{\left( \beta \sqrt{g_{00}}%
\right) ^{2\omega -4}}\right] 2^{E\left( d/2\right) }\left( \omega
-2\right) \left( 1-2^{3-\omega }\right) \left( 3-2^{3-\omega }\right) \left[
\frac{\Gamma \left( \omega -2\right) \zeta \left( \omega -2\right) }{\Gamma
\left( \frac{\omega -1}{2}\right) \left( 2\sqrt{\pi }\right) ^{\omega -1}}%
\right] ^{2}.  \label{eq 2.31}
\end{align}%
From that, we may see the difference between the two contributions: Eqs.\eqref{eq 2.28} and \eqref{eq 2.31}, especially regarding the temperature dependence. Nonetheless, in the same way as happened in the one-loop expression, the higher-derivative contribution manifests the Tolman-Ehrenfest effect, then respecting the scale invariance as well.

It is a long-time well-known fact that higher-loops corrections to the effective Lagrangian in QED may exhibit infrared divergences which arise from the dominant high-temperature contribution of the zero mode. Hence, in order to deal with these divergences, it was proposed a systematic way in dealing to them: one should sum an infinite series of one-particle irreducible diagrams of the one-loop static photon self-energy which are individually divergent. This was also investigated from QED in the background of static gravitational fields \cite{ref13}. And, it was shown that they are equivalent to the pressures of a plasma at high temperature, with the temperature replaced by the local temperature $T_{loc}$. Also, this shows that the Tolman-Ehrenfest effect is explicitly manifested even when
the quantum corrections are taken into account.

Nevertheless, it was shown previously that though the addition of the Podolsky's term (a massive sector) into the Maxwell action enhances the ultraviolet behavior of the whole theory, it does not change the infrared behavior of the radiative correction quantities \cite{ref19} (actually, it is only infrared safe at the known Fried-Yennie gauge $\xi =3$);§ this is because there is still a massless mode propagating. Therefore, we may conclude that this is also true here in the finite temperature case, and the infrared divergences here should be dealt in the same way as it is in the massless theory. This subject is under consideration and is part of a systematic study of nonperturbative phenomena of the generalized electrodynamics at finite temperature.


\section{Concluding Remarks}

\label{sec:3}

In this paper we presented a study on the thermodynamical properties of the generalized electrodynamics in a static gravitational background by evaluating systematically the one-loop and two-loop expressions for the effective Lagrangian. One of the main motivations of the present study was the recent result showing the equivalence between static and zero energy-momentum thermal amplitudes, which holds for the leading contributions at high temperature. This correspondence played an important role in obtaining the effective Lagrangian expressions for static gravitational fields interacting with a plasma of \emph{generalized} photons and massless electrons at high temperature.

It is a remarkable result that, in the same way as it happened in QED, the contributions from the generalized electrodynamics at $\beta M \gg 1$ arising from the approximation of static gravitational fields correspond to those obtained by the ordinary theory defined in Minkowski spacetime, by the simple replacement of an overall factor of $\sqrt{|g|}$ an imprint of a density, and moreover by the modification of the (\emph{global}) Minkowski temperature by the Tolman \emph{local} temperature. Surprisingly, although a scale $M$ has been introduced in the theory, due to the higher-derivative term, the (Weyl) scale invariance is preserved in the whole expression. Nevertheless, since heat interacts with gravity, one may reason naturally that the emergence of a local temperature is unavoidable for a system in a configuration of thermal equilibrium to obviate the heat flowing from regions with different values of gravitational potential \cite{ref14}.

Although we have tried to maintain our treatment as much as arbitrary, especially in which concerns the spacetime dimensions, mainly because of the several studies on unified field theories, we are always bounded by the direct physical application, and they all lie in a four-dimensional spacetime. As aforementioned, a subsequent study of the present analysis will consist in a systematic discussion on nonperturbative phenomena of generalized electrodynamics at finite temperature. Because, though we have massive modes propagating in the gauge field, we still have the presence of the massless modes. And, it is precisely these massless modes that generate the infrared divergences when higher-order contributions are present; such divergence may be traced back to the dominant high-temperature contribution of the zero mode. By means of complementarity, we may also investigate higher-loops contributions \cite{ref26} to realize if the higher-derivative contributions still respects the scale invariance, i.e., if they exhibit a dependence in terms of the Tolman \emph{local} temperature. These issues and others will be further elaborated, investigated and reported elsewhere.

\subsection*{Acknowledgments}

The author would like to thank the anonymous referee for his/her comments and suggestions to improve this paper. R.B. thanks FAPESP for full support.




\begin{thebibliography}{99}

\newcommand{\arXiv}[2]{\href{http://arxiv.org/abs/#1}{arXiv:#1 [#2]}}
\newcommand{\arXivOld}[1]{\href{http://arxiv.org/abs/#1}{arXiv:#1}}		
\bibitem{ref1} E. Braaten and R.D. Pisarski, Nucl. Phys. \textbf{B337}, 569 (1990);  Nucl. Phys. \textbf{B339}, 310 (1990).

\bibitem{ref2} A. Rebhan, Nucl. Phys. \textbf{B351}, 706 (1991).

\bibitem{ref3} J. Frenkel and J.C. Taylor, Nucl. Phys. \textbf{B334}, 199 (1990).

\bibitem{ref4} J.C. Taylor and S.M.H. Wong, Nucl. Phys. \textbf{B346}, 115 (1990).

\bibitem{ref5} F.T Brandt and J. Frenkel, Phys. Rev. D \textbf{47}, 4688 (1993); \arXivOld{hep-ph/9209265}.

\bibitem{ref6} F.T. Brandt, J. Frenkel, and J.C. Taylor, Nucl. Phys. \textbf{B814}, 366 (2009); \arXiv{0901.3458}{hep-th}.

\bibitem{ref7} R.R. Francisco and J. Frenkel, Phys. Lett. B \textbf{722}, 157 (2013); \arXiv{1301.2108}{hep-th}.

\bibitem{ref8} F.T Brandt, J. Frenkel, and J.C. Taylor, Nucl. Phys. \textbf{B437}, 433 (1995); \arXivOld{hep-th/9411130}.

\bibitem{ref9} J. Frenkel, S.H. Pereira, and N. Takahashi, Phys. Rev. D \textbf{79}, 085001 (2009); \arXiv{0902.0757}{hep-th}.

\bibitem{ref10} F.T. Brandt and J.B. Siqueira, Phys. Rev. D \textbf{86}, 105001 (2012); \arXiv{1208.6585}{hep-th}.

\bibitem{ref11} F.T Brandt, J. Frenkel, and J.B. Siqueira, Eur. Phys. J. C \textbf{73}, 2622 (2013); \arXiv{1310.3246}{hep-th}.

\bibitem{ref12} F.T. Brandt and J.B. Siqueira, Phys. Rev. D \textbf{85}, 067701 (2012); \arXiv{1203.1670}{hep-th}.

\bibitem{ref13} F.T Brandt, J. Frenkel, and J.B. Siqueira, Phys. Rev. D \textbf{89}, 045011 (2014); \arXiv{1311.2519}{hep-th}.

\bibitem{ref14} R.C. Tolman, Phys. Rev. \textbf{35}, 904 (1930);

R.C. Tolman and P. Ehrenfest, Phys. Rev. \textbf{36}, 1791 (1930);

Richard C. Tolman, \emph{Relativity, Thermodynamics and Cosmology}, (Dover Publications, New York, 2003). 

\bibitem{ref15} B. Podolsky, Phys. Rev. 62, \textbf{68} (1942);

B. Podolsky and C. Kikuchy, Phys. Rev. \textbf{65}, 228 (1944);

B. Podolsky and P. Schwed, Rev. Mod. Phys. \textbf{20}, 40 (1948).

\bibitem{ref16} C.A.P. Galv\~{a}o and B.M. Pimentel, Can. J. Phys. \textbf{66}, 460 (1988).

\bibitem{ref17} R. R. Cuzinatto, C. A. M. de Melo, and P. J. Pompeia, Ann. Phys. (Amsterdam) \textbf{322}, 1211 (2007); \arXivOld{hep-th/0502052}.

\bibitem{ref18} C. A. Bonin, R. Bufalo, B. M. Pimentel, and G. E. R. Zambrano, Phys. Rev. D \textbf{81}, 025003 (2010); \arXiv{0912.2063}{hep-th};

C. A. Bonin and B. M. Pimentel, Phys. Rev. D \textbf{84}, 065023 (2011); \arXiv{1105.3920}{hep-th}.

\bibitem{ref19} R. Bufalo, B.M. Pimentel, and G.E.R. Zambrano, Phys. Rev. D \textbf{83}, 045007 (2011); \arXiv{1008.3181}{hep-th};

R. Bufalo, B.M. Pimentel, and G.E.R. Zambrano, Phys. Rev. D \textbf{86}, 125023 (2012); \arXiv{1212.3542}{hep-th};

R. Bufalo and B. M. Pimentel, Phys. Rev. D \textbf{88}, 065013 (2013); \arXiv{1404.0940}{hep-th}.

\bibitem{ref20}  R. Bufalo and B.M. Pimentel, Eur. Phys. J. C \textbf{74}, 2993 (2014); \arXiv{1406.2941}{hep-th}.

\bibitem{ref21} J.I. Kapusta and C. Gale, \emph{Finite-Temperature Field Theory}, 2nd ed. (Cambridge University Press, Cambridge, 2011).

\bibitem{ref22} A. Das, \emph{Finite Temperature Field Theory} (World Scientific, New York, 1997).

\bibitem{ref23} N. Birrel and P. Davies, \emph{Quantum Fields in Curved Space},
(Cambridge University Press, Cambridge, 1982).

\bibitem{ref24} M. Abramowitz and I.A. Stegun, \emph{Handbook of Mathematical Functions: With Formulas, Graphs,
and Mathematical Tables}, (Dover, New York, 1965).

\bibitem{ref25} Jean Zinn-Justin, \emph{Quantum Field Theory and Critical Phenomena}, 4th ed. (Oxford University Press, Oxford, 2002).

\bibitem{ref26} C. Corian\'o and R.R. Parwani, Phys. Rev. Lett. \textbf{73}, 2398 (1994);

J.-P. Blaziot, E. Iancu, and R.R. Parwani, Phys. Rev. D \textbf{52}, 2543 (1995).


\end{thebibliography}
\end{document}